\documentclass[conference]{IEEEtran}
\IEEEoverridecommandlockouts
\usepackage{amsmath,amssymb,amsfonts}
\usepackage{amsthm}
\usepackage{bm}
\usepackage{enumitem}
\usepackage{graphicx}
\usepackage{silence}
\usepackage{subcaption}
\usepackage{booktabs}
\usepackage{multirow}
\usepackage{colortbl}
\usepackage{adjustbox}
\usepackage{multicol}
\usepackage{balance}
\usepackage{xcolor}
\usepackage{textcomp}
\usepackage{cite}
\usepackage{url}
\usepackage[most]{tcolorbox}
\tcbuselibrary{theorems, skins, breakable}
\usepackage[linesnumbered,ruled,vlined]{algorithm2e}
\newtcbtheorem{styleddef}{Definition}{
  enhanced,
  breakable,
  colback=gray!12,
  colframe=gray!55,
  boxrule=0.5pt,
  arc=3mm,
  left=2.6mm,right=2.6mm,top=2.2mm,bottom=2mm,
  fonttitle=\bfseries,
  coltitle=white,
  colbacktitle=black!75,
  title filled
}{def}
\definecolor{Gray}{gray}{0.9}
\theoremstyle{definition}

\def\BibTeX{{\rm B\kern-.05em{\sc i\kern-.025em b}\kern-.08em
    T\kern-.1667em\lower.7ex\hbox{E}\kern-.125emX}}

\newcommand{\affmark}[1]{\textsuperscript{\normalfont #1}}
\newcommand{\corrmark}{\textsuperscript{\ensuremath{\dagger}}}

\begin{document}

\title{Controllable Generative Recommendation via Guided Token Refinement}

\author{
    \IEEEauthorblockN{
        Haichao Zhang\affmark{1,2},
        Zhixuan Liang\affmark{3},
        Chong Zhang\affmark{1,2},
        Zixi Chen\affmark{1},
        Wen Wang\affmark{1},
        Jia Wang\affmark{1,2}\corrmark
    }
    \IEEEauthorblockA{
        \affmark{1}Xi'an Jiaotong-Liverpool University
    }
    \IEEEauthorblockA{
        \affmark{2}University of Liverpool
    }
    \IEEEauthorblockA{
        \affmark{3}The Hong Kong Polytechnic University
    }
    
    \IEEEauthorblockA{
        \textbf{Email:}
        haichao.zhang22@student.xjtlu.edu.cn,
        jia.wang02@xjtlu.edu.cn
    }
    \thanks{\ensuremath{^\dagger}\,Corresponding author.}
    \thanks{This study was funded by the National Natural Science
    Foundation of China (Grant No.~72401232), the Natural Science
    Foundation of the Jiangsu Higher Educational Institution of
    China (Grant No.~23KJB520037), the XJTLU Research Development
    Fund (RDF-21-01-053), and Liaoning Liaohe Lab
    (Grant Nos.~LLL24ZZ-02-01 and LLL24ZZ-02-02).}
}
\maketitle

\begin{abstract}
Generative recommendation over discrete Semantic IDs has become a powerful paradigm for item retrieval. Real-world systems, however, frequently require real-time controllability to balance accuracy against objectives such as diversity. Existing autoregressive methods generate tokens strictly from left to right. This fixed direction locks in early decisions, preventing holistic planning and making control difficult. Consequently, direct interventions often break the semantic structure, producing non-existent items. 
To address this, we propose \textbf{ReGen}, a framework that shifts from sequential decoding to \textit{parallel iterative refinement}. ReGen reformulates generation as a \textit{mask-filling} task, predicting all tokens simultaneously rather than one by one. This global view allows us to apply control signals directly to tokens with high-level semantics without retraining the model. 
To ensure that decoded candidates map to real items, we introduce a topology-constrained tree that iteratively corrects low-confidence predictions while strictly enforcing valid codebook paths during beam-based inference.
Experiments show that ReGen provides a favorable accuracy–diversity trade-off under moderate guidance while maintaining high validity of generated Semantic IDs. The source code is available at: \url{https://github.com/zhang-haichao/ReGen}.
\end{abstract}


\section{Introduction}

\begin{figure}[tbp]
    \centering
    \includegraphics[width=1.0\columnwidth]{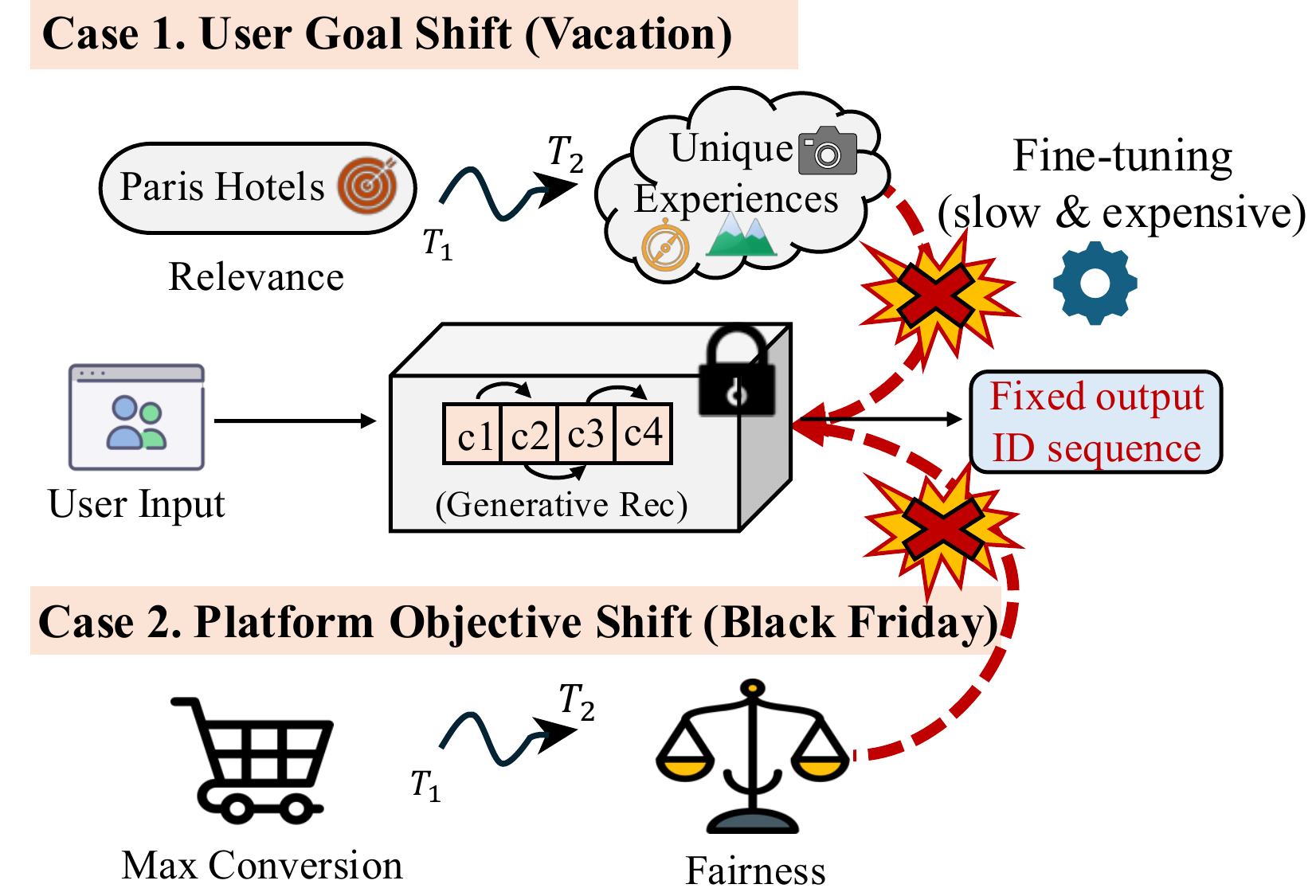}
    \caption{Illustration of rigidity in conventional generative recommendation under dynamic environments. Facing goal shifts (Cases 1 \& 2), autoregressive models suffer from unidirectional decision locking. Fine-tuning is slow and expensive, rendering real-time controllable generation impractical.}
    \label{fig:intro}
     % \vspace{-0.55cm}
\end{figure}

Generative recommendation has emerged as a new paradigm by reformulating item retrieval as sequence generation over discrete semantic identifiers~\cite{petrov2023generative,rajput2023recommender,hou2025generating,deldjoo2024recommendation}. By leveraging advanced generative models, this paradigm enables richer contextual modeling of user behavior than traditional ranking-based approaches~\cite{rajput2023recommender,koren2009matrix,he2020lightgcn}. However, deploying generative recommenders in practice requires not only accurate next-item generation but also the ability to adapt to changing and often conflicting objectives (e.g., relevance vs.\ diversity, utility vs.\ fairness) on the fly~\cite{gogna2017balancing,cheng2017learning}. For example, as shown in Figure~\ref{fig:intro}, a user planning a vacation might initially demand highly relevant results when searching for \textit{Paris hotels}. After browsing multiple options, the user's goal might shift to exploring unique local experiences, immediately prioritizing novelty and diversity. From a platform's perspective, during a limited-time sale like Black Friday, the initial strategy might be to maximize conversion rates. But midway through, the platform might need to quickly adjust its strategy, encouraging the exploration of underrepresented semantic niches or diverse item types to avoid filter bubbles. These rapid objective shifts make static training or repeated fine-tuning impractical, motivating \textbf{r}eal-time \textbf{c}ontrollable \textbf{g}enerative \textbf{r}ecommendation: guiding generation at inference time with adjustable control signals.

Despite extensive research, existing methods struggle to meet this need.
Traditional controllable methods, such as MIND~\cite{li2019multi} and ComiRec~\cite{cen2020controllable}, introduce control through multi-interest modeling but rely on explicit retrieval steps, which do not fit the sequence generation paradigm.
Within generative recommendation, controllability has mainly been introduced in three ways:
(i) Training-time control methods embed control signals into model parameters~\cite{shen2025paragon,he2022metabalance}. However, changing goals requires re-training or fine-tuning,  which is impractical for real-time adjustments. (ii) Post-hoc methods (e.g., re-ranking)~\cite{wang2023diversity} only filter items that were already generated. They cannot create new, diverse items if the generator didn't produce them initially.  (iii) Inference-time control: most models use autoregressive decoding, generating tokens strictly from left to right. This creates \textit{decision locking}: early token choices constrain all future steps~\cite{rajput2023recommender,shi2025llada,chang2022maskgit}. If we try to force control signals mid-generation, it often breaks the semantic structure, producing invalid IDs that do not map to real items~\cite{rajput2023recommender,hou2025generating}.

To address these limitations, we propose \textbf{ReGen}, a real-time controllable generative recommendation framework that enables flexible inference-time control while preserving recommendation accuracy and output validity. Specifically, we propose Hierarchical Logit Guidance, which directly modulates the logits of coarse-grained semantic tokens according to the desired control signal (e.g., semantic diversity) during inference. This lightweight, training-free intervention enables efficient and immediate adjustment of the generation behavior without requiring any modification to the model parameters. However, aggressive logit manipulation may introduce semantic mismatches or predictions that deviate from valid item identifiers. To counteract these risks, ReGen incorporates Confidence-Aware Refinement with Topology-Constrained Tree. This component iteratively identifies low-confidence tokens through a mask predictor, re-predicts them in subsequent refinement steps, and enforces structural validity by restricting all updates to semantically consistent paths within the discrete item topology, ensuring that every generated identifier corresponds to a real, existing item. 
While ReGen can theoretically adapt to various control objectives by formulating specific control signals, we focus on diversity control as a representative and challenging case study in this paper to evaluate its effectiveness.
Our contributions are summarized as follows:
\begin{itemize}
    \item We propose \textbf{ReGen}, a real-time controllable framework based on parallel iterative refinement. It enables global semantic planning and breaks the limitations of autoregressive models.
    \item We introduce Hierarchical Logit Guidance, applying control signals to high-level semantic tokens during inference, enabling flexible control without parameter updates.
    \item We develop Confidence-Aware Refinement with a Topology-Constrained Tree and integrate it into beam-based inference. This mechanism repairs unreliable tokens, preserves valid Semantic ID paths, and produces a compact set of controllable recommendation candidates.
    \item Extensive experiments show that ReGen achieves a favorable accuracy--diversity trade-off under moderate guidance, and ablation studies verify the importance of guidance, refinement, and topology-constrained inference.
\end{itemize}

\section{Related Work}

\subsection{Generative Recommendation System}

Generative recommendation reformulates recommendation as sequence generation, typically using Transformer-based models to predict semantic item ID sequences~\cite{ji2024genrec,wu2024survey,petrov2023generative,hou2025actionpiece}. 
Various Semantic ID construction methods have been explored, including vector quantization with RQ-VAE~\cite{rajput2023recommender,li2025semantic}, structure-aware hierarchical clustering~\cite{si2024generative}, and direct encoding of collaborative signals~\cite{mo2024min}. 
These approaches enable coherent and personalized recommendations by integrating semantic and behavioral information. 
Despite this progress, mainstream generative recommendation models struggle to simultaneously achieve high accuracy, effective diversity control, and efficient inference. 
In particular, diversity control remains challenging under autoregressive generation. 
Methods such as TIGER~\cite{rajput2023recommender} and DLCRec~\cite{chen2025dlcrec} suffer from irreversibility in left-to-right decoding, where prefix perturbations often lead to invalid item IDs and degraded recommendation accuracy. 
These limitations highlight the need for a new generative framework that enables efficient and reliable controllability without sacrificing recommendation quality.

\subsection{Controllable Recommender System}
Controllable recommender systems aim to enhance user autonomy and overcome the limitations of traditional systems that overemphasize accuracy while neglecting flexibility and diversity. 
Early multi-interest models, such as MIND~\cite{li2019multi} and ComiRec~\cite{cen2020controllable}, implicitly or explicitly capture diverse user preferences through dynamic routing and controllable aggregation. 
User-centric methods, including causal UCRS~\cite{tan2023user}, introduce counterfactual reasoning to balance accuracy and diversity. 
Recent LLM-based frameworks provide finer control. 
For instance, DLCRec~\cite{chen2025dlcrec} adopts data-augmented subtask decomposition~\cite{lu2024aligning} and employs supervised learning and RLHF for improved re-ranking and category balance. 
Meanwhile, Paragon~\cite{shen2025paragon} generates conditional parameters for adaptive multi-task learning.
However, LLM-based approaches often suffer from high latency and retraining costs~\cite{11391920}. Conversely, efficiency-oriented methods often compromise real-time adaptability, underscoring the need for a unified framework that achieves controllability and efficiency simultaneously without significant training overhead.

\section{Preliminaries}
\label{sec:preliminaries}

We first introduce the notation for generative recommendation based on Semantic IDs and formalize the inference-time control problem.

\begin{styleddef}{Semantic ID}{semantic_id}
\label{Definition 3.1}
Each item $i$ is represented by a \textbf{Semantic ID} $\mathbf{x}$, defined as a sequence of discrete tokens
$\mathbf{x} = (c^{1}, \dots, c^{L})$, where each $c^{l}$ is an index from a finite codebook.
Semantic IDs are obtained by applying a tokenizer that maps item representations into discrete codes,
such as vector quantization-based methods (e.g., RQ-VAE~\cite{lee2022autoregressive}) or product quantization variants (e.g., OPQ~\cite{ge2013optimized}).
This discrete representation enables generative recommendation models to operate over token sequences rather than continuous embeddings.
\end{styleddef}

\begin{styleddef}{Generative Recommendation Task}{gen_rec_task}
Let $\mathcal{U}$ and $\mathcal{I}$ denote the sets of users and items.
For each user $u \in \mathcal{U}$, we observe a time-ordered interaction history
$\mathcal{S}_u=(i_1,\dots,i_n)$ with $i_k \in \mathcal{I}$. In the generative paradigm, each item $i_k$ is first \textbf{mapped} to its corresponding Semantic ID $\mathbf{x}_k$ (Definition~\ref{def:semantic_id}), converting the interaction history into a semantic sequence $\mathbf{X}_u = (\mathbf{x}_1, \dots, \mathbf{x}_n)$. A generative recommender then models the conditional probability $p_\theta(\mathbf{x} \mid \mathbf{X}_u)$, where $\mathbf{x}$ is the Semantic ID of the target item $i_{n+1}$. The model assigns probabilities to valid Semantic IDs and decodes high-scoring identifiers as recommendation candidates, avoiding exhaustive scoring over the full item catalog.

\end{styleddef}

\noindent\textbf{Problem Statement (Controllable Inference).}
Given a pre-trained backbone $p_\theta(\mathbf{x}\mid \mathbf{X}_u)$ and a control signal $\mathbf{y}$ (e.g., diversity), our goal is to perform \emph{training-free controlled inference}: steering the generation process without updating model parameters. For each valid Semantic ID $\mathbf{x}\in\mathcal{T}$ that maps to an existing item, the controlled generation score is defined as:
\begin{equation}
\Phi_{\mathbf{y}}(\mathbf{x};\mathbf{X}_u,\mathcal{S}_u)
=
\underbrace{\log p_\theta(\mathbf{x} \mid \mathbf{X}_u)}_{\text{Preference Utility}}
+
\lambda
\underbrace{\log p(\mathbf{y}\mid \mathbf{x},\mathcal{S}_u)}_{\text{Controllability}},
\label{eq:problem_statement}
\end{equation}
where $\lambda$ controls the guidance strength, and $p(\mathbf{y}\mid \mathbf{x},\mathcal{S}_u)$ measures the compatibility between candidate $\mathbf{x}$ and the control signal. For diversity, this term penalizes coarse semantic tokens that frequently appear in the user's history $\mathcal{S}_u$.

During inference, ReGen decodes a candidate pool of valid Semantic IDs $\mathcal{X}_u^{\mathrm{cand}} = \operatorname{Top}_{B,\mathbf{x}\in\mathcal{T}} \Phi_{\mathbf{y}}(\mathbf{x};\mathbf{X}_u,\mathcal{S}_u)$ using beam size $B$. Since multiple Semantic IDs can map to the same item, we use the mapping function $\pi(\mathbf{x})$ to deduplicate and filter out historical interactions, yielding a unique item candidate pool $\mathcal{C}_u = \{ \pi(\mathbf{x}) \mid \mathbf{x}\in\mathcal{X}_u^{\mathrm{cand}}, \pi(\mathbf{x})\notin\mathcal{S}_u \}$. For each item $i\in\mathcal{C}_u$, its item-level utility is defined as $\phi_{\mathbf{y}}(i) = \max_{\mathbf{x}\in\mathcal{X}_u^{\mathrm{cand}}, \pi(\mathbf{x})=i} \Phi_{\mathbf{y}}(\mathbf{x};\mathbf{X}_u,\mathcal{S}_u)$.

Finally, the top-$k$ recommendation list $\mathcal{R}_u$ is constructed by greedily selecting items from $\mathcal{C}_u$ that maximize both utility and list-level controllability:
\begin{equation}
\mathcal{R}_u = \operatorname{Select}_{k} (\mathcal{C}_u;\phi_{\mathbf{y}}, C_{\mathbf{y}}).
\label{eq:ranked_output}
\end{equation}
In this work, we concretely instantiate the list-level control criterion $C_{\mathbf{y}}$ as category coverage $C_{\mathbf{y}}(\mathcal{R}_u) = \left| \{ g(i) \mid i \in \mathcal{R}_u \} \right|$ (where $g(i)$ denotes the category label of item $i$), with its marginal gain evaluated during selection defined as:
$$
\Delta C_{\mathbf{y}}(i \mid \mathcal{R}_u) = C_{\mathbf{y}}(\mathcal{R}_u \cup \{i\}) - C_{\mathbf{y}}(\mathcal{R}_u).
$$
The final output thus strictly satisfies preference utility, targeted controllability, and semantic validity.

\section{Methodology}
\label{sec:methodology}

\begin{figure*}[t]
    \centering
    % \vspace{10pt}
    \includegraphics[width=2\columnwidth]{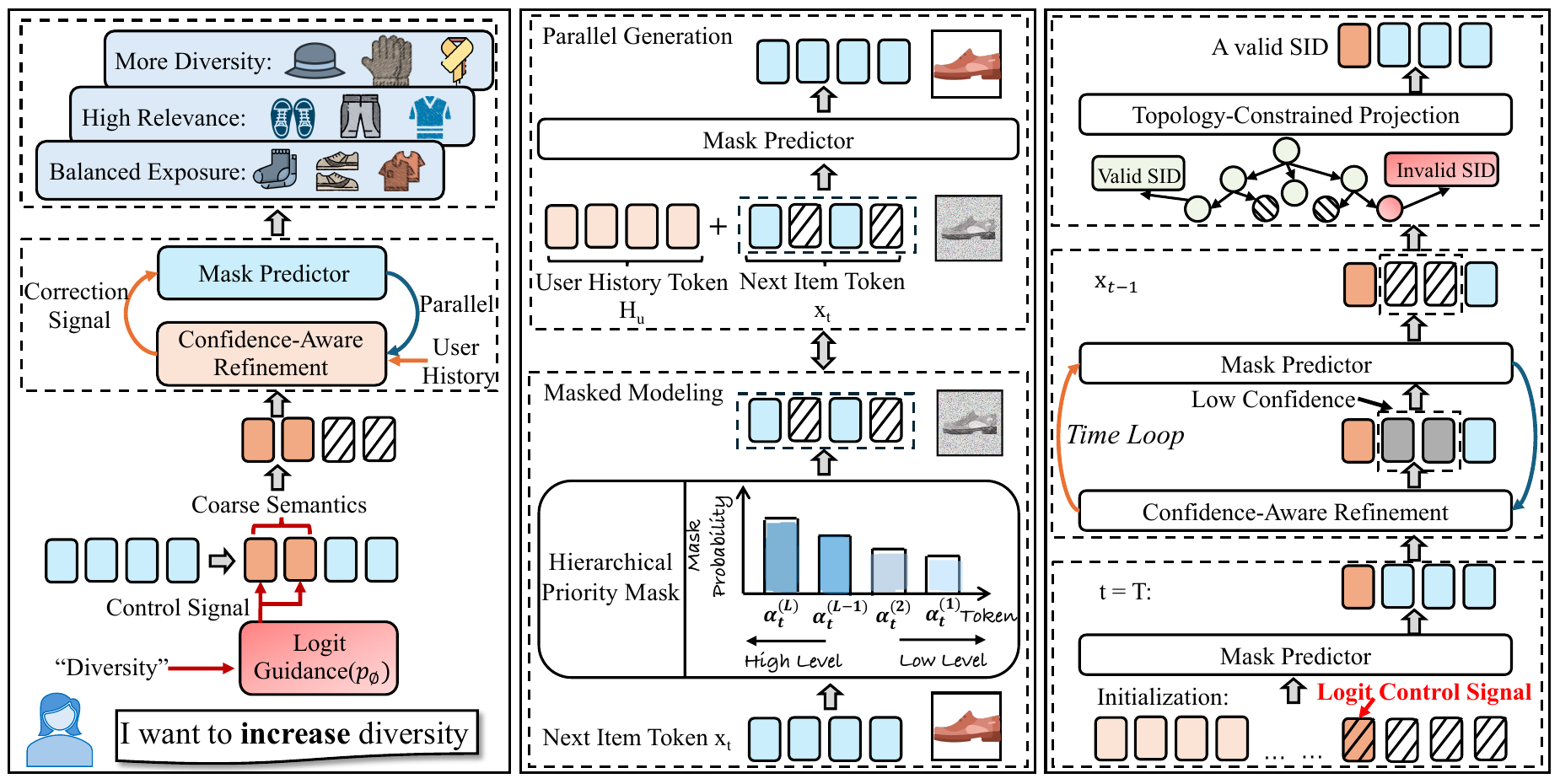}
    \caption{Overview of the ReGen framework. ReGen achieves real-time controllable recommendation through the following phases: (1) \textbf{Controllable Inference Workflow} (Left): Illustrates how control signals guide the coarse-grained semantics during inference. (2) \textbf{Parallel Semantic ID Generation} (Middle): The backbone model captures global context for parallel prediction. (3) \textbf{Confidence-Aware Refinement} (Right): An iterative loop to ensure coherence and validity.}
    \label{fig:frame}
    % \vspace{-0.35cm}
\end{figure*}

We propose ReGen, a framework achieving Real-time Controllable Generative Recommendation via parallel iterative refinement.
As shown in Figure~\ref{fig:frame}, ReGen consists of three core modeling components together with a lightweight beam-based inference procedure:
(1) \textbf{Parallel Semantic ID Generation:} we build a parallel generative model that predicts all Semantic ID tokens jointly and updates them iteratively. It captures global context and supports direct updates of coarse-grained tokens, which prepares the model for inference-time control.
(2) \textbf{Logit Control Signal:} we apply the control signal to the logits of selected coarse-grained tokens during inference. This guides the generation toward desired attributes without updating model parameters, while keeping the intervention lightweight.
(3) \textbf{Confidence-Aware Refinement:} to recover quality under control, we identify low-confidence tokens and regenerate them in the iterative loop. We also enforce structural constraints so decoded Semantic IDs remain valid and map to real items.
During inference, these components are executed with a lightweight beam mechanism that retains multiple valid Semantic ID candidates, maps them to items after deduplication, and constructs the final ranked output within this generated candidate pool.

\subsection{Parallel Semantic ID Generation}
\label{sec:parallel_generation}

In this section, we introduce a framework to facilitate real-time controllability. 
Unlike traditional autoregressive models that generate Semantic IDs token by token, we reformulate the process as a parallel fill-in-the-blank task. 
By leveraging global bidirectional context, our model can jointly predict and refine tokens, ensuring that high-level control signals are seamlessly integrated with fine-grained item details.

\subsubsection{Semantic ID Tokenization}
\label{sec:tokenization}

To enable generative recommendation, we first map items into a discrete semantic space.
We use residual vector quantization (RQ-VAE)~\cite{lee2022autoregressive,rajput2023recommender} to encode each item $i$ as a sequence of discrete tokens
$\mathbf{x} = (c^{1}, \dots, c^{L})$, where $L$ is the code length.
The item embedding $\mathbf{e}_i$ is approximated as a sum of code embeddings,
\begin{equation}
\mathbf{e}_i \approx \sum_{l=1}^{L} \mathbf{e}_{c^l}, \quad \mathbf{e}_{c^l} \in \mathcal{C}_l,
\end{equation}
where $\mathcal{C}_l$ denotes the codebook at level $l$.
This quantization induces a coarse-to-fine organization, with earlier tokens capturing higher-level semantics and later tokens encoding finer details.
We leverage this hierarchy as a structural semantic prior for control. 
Our parallel generative framework uses higher-level tokens as semantic anchors for controllable intervention, while updating the remaining tokens in parallel to preserve coherence and validity.

\subsubsection{Parallel Generative Model}
\label{sec:parallel_structure}

We implement the parallel generative model $f_\theta$ using a bidirectional Transformer encoder~\cite{vaswani2017attention}, following a BERT-style masked modeling architecture~\cite{devlin2019bert}.
Formally, the backbone takes two inputs: the semantic history $\mathbf{X}_u$ and a candidate Semantic ID sequence $\mathbf{x} = (c^{1}, \dots, c^{L})$.
To facilitate parallel generation, the candidate sequence $\mathbf{x}$ is composed of a mixture of determined tokens (e.g., fixed control signals) and the special mask symbol \texttt{[MASK]}.
The model processes these inputs simultaneously to produce a sequence of logit vectors for all $L$ positions in parallel:
\begin{equation}
[\mathbf{Z}^{(1)}, \dots, \mathbf{Z}^{(L)}] = f_\theta(\mathbf{x}, \mathbf{X}_u),
\end{equation}
where each $\mathbf{Z}^{(l)} \in \mathbb{R}^{|\mathcal{V}|}$ denotes the unnormalized logit vector over the vocabulary $\mathcal{V}$ at semantic level $l$.

Unlike autoregressive decoding, this architecture treats generation as a global \textit{fill-in-the-blank} problem: each token prediction conditions on the full sequence context via bidirectional attention, rather than only a left prefix.
This property is crucial for controllable recommendation: we can fix coarse-grained tokens to reflect the control signal (e.g., diversity) and let the model update the remaining fine-grained tokens under the new global semantics, without regenerating the entire sequence from scratch.

\subsubsection{Masked Predictor}
\label{sec:Masked_Predictor}
The Mask Predictor (shown in Figure~\ref{fig:frame}) is the prediction head of our parallel generative model. It predicts the tokens at masked positions and is used to complete or update a Semantic ID. Concretely, the backbone $f_\theta$ outputs unnormalized logits, and the Mask Predictor applies softmax to obtain token probabilities.
Let $\mathbf{x}=(c^{1},\dots,c^{L})$ be a partially specified Semantic ID. We use $M \subseteq \{1,\dots,L\}$ to denote masked positions and $\bar{M}=\{1,\dots,L\}\setminus M$ to denote observed positions. The observed tokens can include control-fixed tokens, while the remaining positions are masked with \texttt{[MASK]}.

Given $(\mathbf{x}_{\bar{M}},\mathbf{X}_u)$, the backbone produces a logit vector $\mathbf{Z}^{(l)} \in \mathbb{R}^{|\mathcal{V}|}$ for each level $l$. The Mask Predictor converts it into a token distribution: 
\begin{equation} 
p_\theta\!\left(c^{l} \mid \mathbf{x}_{\bar{M}}, \mathbf{X}_u\right) = \mathrm{Softmax}\!\left(\mathbf{Z}^{(l)}\right), \qquad l = 1, \dots, L, 
\end{equation}
where $\mathcal{V}$ is the token vocabulary. We compute these distributions for all masked positions $l\in M$ in parallel, and then fill or expand the masked tokens during inference to obtain updated Semantic ID candidates.

This module provides a direct interface for controllable inference. When we modify a few coarse-grained tokens to satisfy a diversity control signal, we keep these tokens in the observed set $\bar{M}$. The Mask Predictor then re-predicts the remaining masked tokens conditioned on the modified high-level semantics and the user history, which helps keep the completed Semantic ID coherent.

% \textbf{Hierarchical priority masking.} 
To support coarse-to-fine control, we employ a \textit{hierarchical priority masking} strategy during training. We assign lower masking probabilities to coarse-grained tokens ($\alpha_1 < \cdots < \alpha_L$). This forces the model to learn to reconstruct fine-grained details conditioned on stable high-level semantics.

\subsection{Logit Control Signal}

\label{sec:logit_control_signal}

In this subsection, we introduce a Logit Control Signal to enable real-time diversity control in generative recommendation. 
The key idea is to directly intervene on the output logits at inference time, so the model becomes more likely to produce items with desired attributes (e.g., categories the user has not recently seen). 
Importantly, this is \emph{training-free}: we do not update any model parameters. 
However, a practical challenge remains: how can we reliably impose such diversity constraints on a multi-level Semantic ID?

\subsubsection{Hierarchical Logit Guidance}
Inspired by the hierarchical codebooks in RQ-VAE, we recognize that global semantics are determined by the top-level codes, which implicitly capture macro-level item attributes such as categories. Therefore, we apply the control signal to coarse-grained tokens. This macro-level semantic guidance naturally translates into categorical diversity in the item space, while we rely on the parallel generator to update the remaining fine-grained tokens accordingly.
In ReGen, the compatibility term in Eq.~\eqref{eq:problem_statement} is instantiated through these level-aware control scores and injected directly into the decoding logits.

To implement diversity control, we convert the requirement
into level-aware control score vectors
$\{\mathbf{s}_{\mathbf{y}}^{(l)}\}_{l=1}^{K}$ over coarse-grained
Semantic ID levels. We formulate this score as a frequency-based
penalty to suppress semantic regions that appear frequently in the
user's history $\mathcal{S}_u$:
\begin{equation}
\mathbf{s}_{\mathbf{y}}^{(l)}[v]
=
-\log\!\left(1+\mathrm{Count}_{l}(v,\mathcal{S}_u)\right),
\quad l \le K,
\end{equation}
where $\mathrm{Count}_{l}(v,\mathcal{S}_u)$ counts how often token
$v$ appears at semantic level $l$ in the user's history. This assigns
higher scores to rare or unseen coarse tokens, explicitly encouraging
exploration in underrepresented semantic regions.

Here, $K \le L$ denotes the number of coarse-grained Semantic ID levels directly controlled during inference.
Let $\mathbf{Z}^{(l)} \in \mathbb{R}^{|\mathcal{V}|}$ denote the
level-$l$ logit vector produced by the backbone $f_\theta$. We inject
the control score into the logits, but only at the first $K$ coarse-grained levels:
\begin{equation}
\tilde{\mathbf{Z}}^{(l)}
=
\begin{cases}
\mathbf{Z}^{(l)} + \lambda \mathbf{s}_{\mathbf{y}}^{(l)}, & l \le K,\\
\mathbf{Z}^{(l)}, & l > K,
\end{cases}
\label{eq:hier_logit_guidance}
\end{equation}
where $\lambda \ge 0$ controls the guidance strength. After applying
softmax to $\tilde{\mathbf{Z}}^{(l)}$, the model becomes more likely
to select coarse tokens that satisfy the control signal.

Applying guidance only on coarse levels exploits the hierarchy of Semantic IDs: coarse tokens set the global semantics, while fine-grained tokens specify item details. After a guided coarse token is selected and kept as context, the Mask Predictor (Section~\ref{sec:Masked_Predictor}) re-predicts the remaining fine-grained tokens without directly biasing them. This keeps the semantic direction aligned with the control signal while preserving sufficient flexibility in deeper levels to maintain recommendation utility.

\subsection{Confidence-Aware Token Refinement}

\label{sec:refinement}
In this subsection, we propose Confidence-Aware Refinement to improve recommendation quality after diversity control. 
While Hierarchical Logit Guidance enforces diversity by changing coarse-grained tokens, it can also create a mismatch between the modified global semantics and the remaining fine-grained tokens, which hurts coherence and relevance.
To address this issue, we use token confidence (e.g., the predicted probability of the selected token) to identify unreliable positions and iteratively re-predict them, while keeping the guided coarse tokens fixed.

% \subsubsection{Iterative Token Refinement}

To resolve the semantic mismatch introduced by diversity guidance, we perform iterative token refinement using the Mask Predictor (Section~\ref{sec:Masked_Predictor}). We run refinement for several rounds; we index these rounds by $T$ and use $\mathbf{x}_t$ to denote the Semantic ID after the $T$-th round. In each round, we keep reliable tokens as context and re-mask and re-predict the rest.

We measure this confidence using the guided logits $\tilde{\mathbf{Z}}_t$ (Eq.~\ref{eq:hier_logit_guidance}), which reflect the shifted distribution after applying the control signal. For each position $l$ at refinement step $T$, the confidence score $\rho_t^{(l)}$ is defined as the maximum softmax probability:

\begin{equation} \rho_t^{(l)} = \max \mathrm{Softmax}(\tilde{\mathbf{Z}}_t^{(l)}). \label{eq:confidence-compute} 
\end{equation}

Under this definition, guided coarse tokens typically have high confidence, whereas fine-grained tokens that no longer fit the new coarse semantics tend to have lower confidence.

Based on these scores, we dynamically decide which tokens to refine. We use a step-dependent threshold $\tau_t$, which typically increases as the refinement progresses. We construct a binary mask vector $\mathbf{m}_t$ where $\mathbf{m}_t^{(l)}=1$ indicates the token needs to be refined:

\begin{equation} 
\mathbf{m}_t^{(l)} = \begin{cases} 1, & \text{if } \rho_t^{(l)} < \tau_t \quad (\text{Low Confidence}), \\ 0, & \text{otherwise} \quad (\text{High Confidence}). \end{cases} 
\end{equation}

Tokens with high confidence ($\mathbf{m}_t^{(l)}=0$) are preserved as anchors. Low-confidence tokens are re-masked with the special symbol \texttt{[MASK]}. We then feed this partially masked sequence and the user history back into the Mask Predictor to generate the next state $\mathbf{x}_{t+1}$: 
\begin{equation} 
\mathbf{x}_{t+1} \leftarrow \operatorname*{arg\,max}_{\mathbf{x} \in \mathcal{V}^{L}} \; p_\theta\left(\mathbf{x} \mid \mathbf{x}_t \odot (1-\mathbf{m}_t) + \texttt{[MASK]} \odot \mathbf{m}_t, \mathbf{X}_u\right).
\end{equation}
The resulting confidence mask determines which positions
are updated during the validity-preserving controlled inference
described in Section~\ref{sec:controlled_inference}.
Across refinement steps, reliable tokens are retained as anchors,
while low-confidence tokens are re-masked and regenerated to
better align the completed Semantic ID with the guided coarse
semantics, thereby improving coherence and recommendation utility
under diversity control.

\subsection{Topology-Constrained Beam Inference}
\label{sec:controlled_inference}
Confidence-aware refinement determines which token positions
should be updated, but parallel updates may still produce
Semantic IDs that do not correspond to catalog items.
We therefore combine topology-constrained projection with
beam-based refinement to search over multiple valid candidates.
The resulting Semantic IDs are subsequently mapped to items
and ranked under the list-level control objective.

\subsubsection{Topology-Constrained Projection}
\label{sec:topology_projection}

Parallel token prediction may produce code combinations that
do not correspond to catalog items. To ensure valid generation,
we organize all catalog Semantic IDs in a prefix tree (Trie)
$\mathcal{T}$ and apply trie-based constrained decoding
\cite{fredkin1960trie,decao2021autoregressiveentityretrieval}.
Each root-to-leaf path in $\mathcal{T}$ represents a valid
Semantic ID.

For a candidate $\mathbf{x}_t$ and a masked position $l$,
let $\mathcal{A}_{\mathcal{T}}(\mathbf{x}_t,l)$ denote the set
of token values that remain compatible with at least one valid
path in $\mathcal{T}$, given the tokens currently fixed in
$\mathbf{x}_t$. Before beam expansion, we suppress every
incompatible token value by
\begin{equation}
\tilde{\mathbf{Z}}_{t}^{(l)}[v]
\leftarrow -\infty,
\qquad
\forall v\notin
\mathcal{A}_{\mathcal{T}}(\mathbf{x}_t,l).
\label{eq:topology_projection}
\end{equation}
Here, $\tilde{\mathbf{Z}}_{t}^{(l)}[v]$ denotes the guided
logit of token $v$ at position $l$. Logits of valid tokens
remain unchanged, so beam expansion considers only token
choices that can be completed into valid catalog Semantic IDs.

\subsubsection{Beam Search and Item Selection}
\label{sec:beam_refinement}

During inference, ReGen retains up to $B$ candidate sequences
at each refinement step. Decoding starts from an all-mask template,
$\mathbf{x}_0=(\texttt{[MASK]},\ldots,\texttt{[MASK]})$.
For each candidate, the confidence mask identifies the positions
to be regenerated. The Mask Predictor predicts these positions,
hierarchical guidance adjusts their logits according to the
control signal, and Eq.~\eqref{eq:topology_projection} removes
token choices that cannot lead to valid catalog IDs. The resulting
candidates are ranked by the accumulated log-probabilities of
the token choices made during refinement, and only the top-$B$
candidates are retained for the next step.

After $T$ refinement steps, the surviving valid Semantic IDs
form the candidate set $\mathcal{X}_u^{\mathrm{cand}}$ for
subsequent item selection. We map each Semantic ID $\mathbf{x}$
to its corresponding item $\pi(\mathbf{x})$, remove items in the
user history, and merge duplicate items to obtain
$\mathcal{C}_u$. When multiple Semantic IDs map to the same item,
its utility $\phi_{\mathbf{y}}(i)$ is the highest controlled score
$\Phi_{\mathbf{y}}$ among the associated IDs, as defined in
Eq.~\eqref{eq:problem_statement}.

We implement the list construction in Eq.~\eqref{eq:ranked_output}
greedily. Starting from $\mathcal{R}_u=\emptyset$, ReGen selects
\begin{equation}
i^\star =
\operatorname*{arg\,max}_{i\in
\mathcal{C}_u\setminus\mathcal{R}_u}
\left[
\phi_{\mathbf{y}}(i)
+
\mu\,\Delta C_{\mathbf{y}}(i\mid\mathcal{R}_u)
\right],
\label{eq:list_selection}
\end{equation}
where $\Delta C_{\mathbf{y}}(i\mid\mathcal{R}_u)$ is the marginal
gain under the list-level control objective, and $\mu\geq 0$
controls its contribution to item selection. The selected item
is added to $\mathcal{R}_u$, and the process is repeated until
$k$ items are obtained. For the diversity-control setting
evaluated in this work, $C_{\mathbf{y}}$ is instantiated as
category coverage.

\section{Experiments}
To evaluate the proposed framework, our experiments focus on four key aspects: (1) the overall recommendation performance compared to SOTA baselines; (2) the controllability and accuracy-diversity trade-off (Pareto frontier); (3) the contribution of core components like guidance and refinement; and (4) the parameter robustness and sensitivity.

\subsection{Experiment Settings}

\noindent \paragraph{Datasets}
We evaluate ReGen on three categories from the Amazon 2023 Review dataset~\cite{hou2024bridging}~\footnote{\url{https://cseweb.ucsd.edu/~jmcauley/datasets/amazon/links.html}}: 
Industrial \& Scientific (Scientific), Musical Instruments (Instrument), and Video Games (Game).
Following common practice~\cite{hou2023learning,rajput2023recommender}, user reviews are treated as interaction sequences and ordered chronologically.
We adopt the leave-one-out protocol~\cite{8594844}, where the last item is used for testing and the second-to-last for validation.
Dataset statistics are summarized in Table~\ref{tab:dataset}.
% \subsubsection{Datasets} 
% We evaluate our method on three categories from the widely used Amazon 2023 Review dataset~\cite{hou2024bridging}: “Industrial \& Scientific” \textbf{(Scientific)}, “Musical \& Instruments” \textbf{(Instrument)}, and “Video \& Games” \textbf{(Game)}. In line with~\cite{hou2023learning,rajput2023recommender}, each user’s historical reviews are treated as interaction records and ordered chronologically, with the earliest review placed first. For evaluation, we adopt the widely used leave-one-out protocol~\cite{kang2018self,rajput2023recommender}: in each user sequence, the last item is held out for testing, and the second-to-last item is reserved for validation. The detailed statistics of the three datasets are presented in Table~\ref{tab:dataset}.
\begin{table}[ht]
\centering
\caption{Statistics of datasets.}
\label{tab:dataset}
\small
\begin{tabular}{lrrrr}
\toprule
Dataset & \#Users & \#Items & \#Interaction & Sparsity \\
\midrule
Instrument & 57,439 & 24,587 & 511,836 & 99.9638\% \\
Scientific  & 50,985 & 25,848 & 412,947 & 99.9687\% \\
Video Games & 94,762 & 25,612 & 814,586 & 99.9664\% \\
\bottomrule
\end{tabular}
% \vspace{-0.2cm}
\end{table}

\begin{table*}[t]
% \vspace{0.3cm}
\centering
\caption{Overall accuracy comparison on three datasets under the accuracy-oriented configuration. The best results are highlighted in \textbf{bold}, and the second-best are \underline{underlined}. For ReGen, we report the configuration selected by validation accuracy to evaluate its standard top-$K$ recommendation utility. For controllable baselines, we report their default accuracy-oriented results without emphasizing additional diversity constraints. This table is intended to compare standard recommendation utility; controllability and accuracy--diversity trade-offs are analyzed separately in Section~\ref{sec:Accuracy-Diversity Trade-off Analysis}.}
\label{tab:masked_results}
% \vspace{0.3cm}
\small
% 1. 将 booktabs 产生的额外垂直留白清零，消除白色断层
% \setlength{\aboverulesep}{0pt}
% \setlength{\belowrulesep}{0pt}
% 2. 因为去掉了留白，表格会变挤，所以稍微加大行高 (1.1 或 1.2 均可)
\renewcommand{\arraystretch}{1.1}
% \setlength{\tabcolsep}{3.5pt}
% 需要在导言区添加: \usepackage{colortbl} 和 \definecolor{Gray}{gray}{0.9}
\adjustbox{max width=1\textwidth}{
\begin{tabular}{cl ccc ccc ccc c}
\toprule
\multirow{2}{*}{Dataset} & \multirow{2}{*}{Metric} 
& \multicolumn{3}{c}{Discriminative Models} 
& \multicolumn{3}{c}{Generative Models} 
& \multicolumn{3}{c}{Controllable Methods} 
& \\ % <--- 【修改点 1】第一行这里留空，千万不要写文字！
\cmidrule(lr){3-5} \cmidrule(lr){6-8} \cmidrule(lr){9-11}
 & 
 & Bert4Rec & SASRec & GRU4Rec
 & VQ-Rec & RPG & LLaDA-Rec
 & ComiRec & Paragon & TIGER-MMR
 & \multirow{-2}{*}{\textbf{ReGen}} \\ % <--- 【修改点 2】把文字移到第二行，并改成 -2 (表示向上跨两行)
\midrule

\multirow{4}{*}{Instruments}
& Recall@5  & 0.0258 & 0.0332 & 0.0310 & 0.0295 & 0.0362 & 0.0347 & 0.0230 & \textbf{0.0396} & 0.0315 & \underline{0.0367} \\
& NDCG@5    & 0.0166 & 0.0177 & 0.0199 & 0.0189 & \textbf{0.0241} & \underline{0.0228} & 0.0150 & 0.0185 & 0.0206 & 0.0219 \\
& Recall@10 & 0.0415 & 0.0545 & 0.0496 & 0.0456 & 0.0545 & 0.0535 & 0.0366 & \textbf{0.0861} & 0.0496 & \underline{0.0646} \\
& NDCG@10   & 0.0216 & 0.0245 & 0.0259 & 0.0240 & 0.0300 & 0.0286 & 0.0194 & \textbf{0.0327} & 0.0264 & \underline{0.0309} \\

\midrule

\multirow{4}{*}{Scientific}
& Recall@5  & 0.0134 & 0.0248 & 0.0190 & 0.0212 & 0.0257 & \underline{0.0265} & 0.0196 & \textbf{0.0363} & 0.0233 & 0.0262 \\
& NDCG@5    & 0.0086 & 0.0133 & 0.0118 & 0.0138 & \underline{0.0174} & 0.0173 & 0.0133 & \textbf{0.0191} & 0.0150 & 0.0154 \\
& Recall@10 & 0.0221 & 0.0392 & 0.0311 & 0.0329 & 0.0395 & 0.0407 & 0.0300 & \textbf{0.0715} & 0.0374 & \underline{0.0491} \\
& NDCG@10   & 0.0114 & 0.0180 & 0.0157 & 0.0175 & 0.0218 & 0.0218 & 0.0167 & \textbf{0.0313} & 0.0195 & \underline{0.0228} \\

\midrule 

\multirow{4}{*}{Video Games}
& Recall@5  & 0.0412 & 0.0575 & 0.0479 & 0.0472 & \underline{0.0579} & \textbf{0.0592} & 0.0325 & 0.0503 & 0.0410 & 0.0400 \\
& NDCG@5    & 0.0263 & 0.0322 & 0.0307 & 0.0309 & \textbf{0.0397} & \underline{0.0394} & 0.0214 & 0.0248 & 0.0269 & 0.0235 \\
& Recall@10 & 0.0667 & \textbf{0.0916} & 0.0776 & 0.0731 & 0.0853 & \underline{0.0895} & 0.0512 & 0.0842 & 0.0648 & 0.0726 \\
& NDCG@10   & 0.0345 & 0.0432 & 0.0402 & 0.0391 & \underline{0.0485} & \textbf{0.0491} & 0.0274 & 0.0381 & 0.0345 & 0.0340 \\
\bottomrule
\end{tabular}
}
\end{table*}
\noindent \paragraph{Baseline Models}
% We compare ReGen with representative discriminative and generative recommenders.
% Discriminative baselines include GRU4Rec~\cite{hidasi2015session}, SASRec~\cite{8594844}, and BERT4Rec~\cite{sun2019bert4rec}.
% Generative baselines based on Semantic IDs include VQ-Rec~\cite{hou2023learning}, RPG~\cite{hou2025generating}, and LLaDA-Rec~\cite{shi2025llada}.
% We also compare with controllable recommendation methods, including ComiRec~\cite{cen2020controllable}, Paragon~\cite{shen2025paragon}, and TIGER-MMR~\cite{rajput2023recommender,carbonell1998use}.

We first compared our approach with discriminative methods based on Item ID: (1) \textbf{GRU4Rec}~\cite{hidasi2015session} leverages gated recurrent units (GRUs) to model user interaction histories; (2) \textbf{SASRec}~\cite{kang2018self} employs a unidirectional Transformer to capture sequential dependencies; (3) \textbf{BERT4Rec}~\cite{sun2019bert4rec} utilizes a bidirectional Transformer trained with a cloze-style objective; We also compared our approach with generative recommendation methods using semantic IDs: (4) \textbf{VQ-Rec}~\cite{hou2023learning} tokenizes items into semantic IDs via product quantization, then pools them for representations. (5) \textbf{RPG}~\cite{hou2025generating} is a lightweight semantic ID–based model that generates long, unordered semantic IDs in parallel using multi-token prediction; (6) \textbf{LLaDA-Rec}~\cite{shi2025llada}, a discrete diffusion-based generative recommender with parallel semantic ID generation and bidirectional modeling. Finally, compared with some Controllable Methods: (7) \textbf{ComiRec}~\cite{cen2020controllable} captures users' multiple interests and generates diverse recommendation lists through dynamic routing and attention mechanism; (8) \textbf{Paragon}~\cite{shen2025paragon} achieves efficient sharing and personalized control by generating task-specific parameters. (9) \textbf{TIGER-MMR} is a TIGER framework~\cite{rajput2023recommender} combined with the maximum marginal relevance mechanism (MMR), which improves the diversity and relevance of recommendation results.

% \noindent \paragraph{Implementation Details}
% For the parallel tokenization module, we adopt Sentence-T5 ~\cite{ni2022sentence} to encode the title and other textual information of each item into an embedding. We employ $M = 4$ codebooks, each containing $K = 256$ code vectors of dimension $D = 32$. 
% The Multi-Head VQ-VAE is trained for 2000 epochs with the AdamW optimizer~\cite{loshchilov2017decoupled}, a learning rate of 1e-3, and a batch size of 2,048. In the discrete diffusion model, the mask predictor is a bidirectional Transformer encoder with a token-embedding dimension of 256 and 8 attention heads per layer. For the Scientific and Instrument datasets, we use a 4-layer encoder; for the Video Game dataset, we use a 6-layer encoder. The model parameters are randomly initialized and trained using our designed loss function. We train the Transformer encoder for 100 epochs with early stopping, using the AdamW optimizer. The learning rate is tuned over \{0.0001, 0.0005, 0.001\} and the weight decay over \{0.05,0.005,0.001\}. The batch size is set to 1,024. The source code is available at: \url{https://anonymous.4open.science/r/ReGen}.

\noindent \paragraph{Evaluation Metrics}
Following prior studies~\cite{tsai2020effects,parapar2021towards,carraro2025enhancing}, we evaluate performance using three commonly adopted ranking metrics: Diversity@$k$, NDCG@$k$, and Recall@$k$. We report results for $k \in \{5, 10\}$. To align with standard industry practices and provide a fair comparison with prior controllable methods (e.g., ComiRec, TIGER-MMR), we evaluate Diversity@$k$ via category coverage. We adopt this as a simple and commonly used proxy metric for global semantic divergence; since our coarse-grained Semantic IDs implicitly encapsulate macro-level item attributes, an increase in category coverage effectively reflects the successful exploration of diverse semantic regions. NDCG@$k$ is used to evaluate the ranking quality of the generated recommendations. Recall@$k$ measures the proportion of relevant ground-truth items that are successfully retrieved in the top-$k$ results.

\subsection{Detailed Implementation}

\paragraph{Semantic Tokenization}
We first construct discrete semantic identifiers for items through a two-stage encoding-and-quantization pipeline. Specifically, we use Sentence-T5~\cite{ni2022sentence} to encode the title and textual metadata of each item into a dense semantic representation. The resulting embeddings are then discretized by an RQ-VAE, which contains $L=4$ residual codebooks. Each codebook consists of $C=256$ code vectors with dimensionality $D=32$, yielding a four-level semantic ID for each item. The RQ-VAE is optimized for 2,000 epochs using AdamW~\cite{loshchilov2017decoupled} with a learning rate of $10^{-3}$ and a batch size of 2,048.

\paragraph{Generative Backbone}
The mask predictor in ReGen is instantiated as a bidirectional Transformer encoder following a masked semantic ID modeling objective. The encoder uses a token embedding dimension of 256 and 8 attention heads in each layer. To accommodate different dataset scales and complexities, we use a 4-layer encoder for the Scientific and Instrument datasets, and a 6-layer encoder for the Video Game dataset. All model parameters are randomly initialized and trained from scratch. The Transformer encoder is trained for up to 100 epochs with early stopping, using AdamW as the optimizer and a fixed batch size of 1,024. Hyperparameters are selected via grid search over learning rates ${10^{-4}, 5 \times 10^{-4}, 10^{-3}}$ and weight decay values ${0.05, 0.005, 0.001}$.

\paragraph{Inference and Ranking Construction}
During inference, ReGen uses the beam-based decoding procedure described in Section~\ref{sec:beam_refinement}. Unless otherwise specified, we set the beam size to $B=50$ and keep the top-$H=100$ valid token expansions at each step. 
After $T$ refinement steps, completed valid Semantic IDs are mapped to items, deduplicated at the item level, and filtered against the user history. The final ranked output is then constructed within the generated candidate pool using the list-level selection rule in Eq.~\eqref{eq:ranked_output}. 
All reported Recall@$k$, NDCG@$k$, and Diversity@$k$ are computed on this final ranked output.

\begin{figure}[t]
    \centering
    % 子图 (a)
    \begin{subfigure}[b]{0.48\textwidth}
        \centering
        \includegraphics[width=\textwidth]{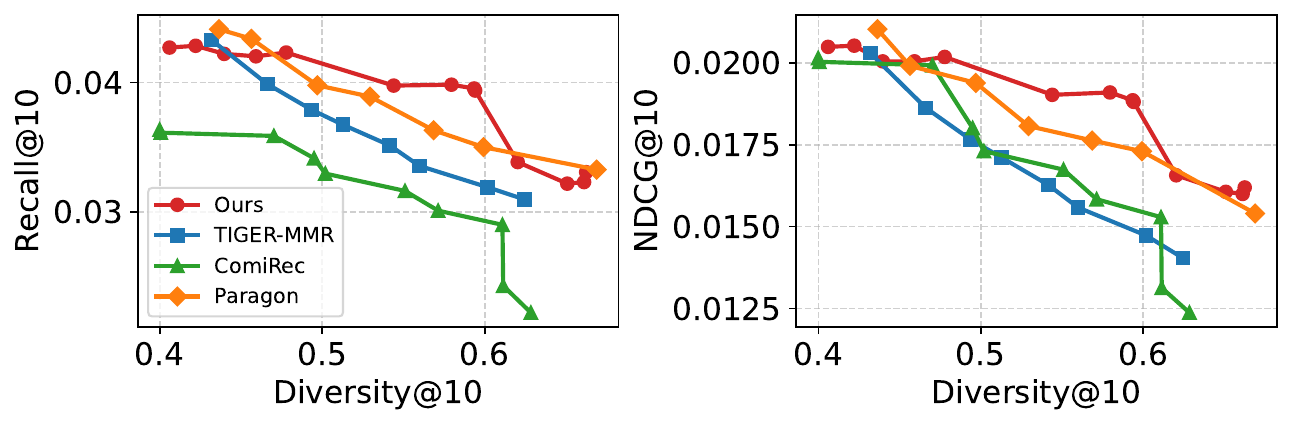}
        % \vspace{-0.6cm}
        \caption{\textbf{Instruments}} % 使用加粗，视觉上更有分量，不显单调
        \label{fig:pareto_instruments}
    \end{subfigure}
    \hfill
    % \vspace{-0.3cm}
    % 子图 (b)
    \begin{subfigure}[b]{0.48\textwidth}
        \centering
        \includegraphics[width=\textwidth]{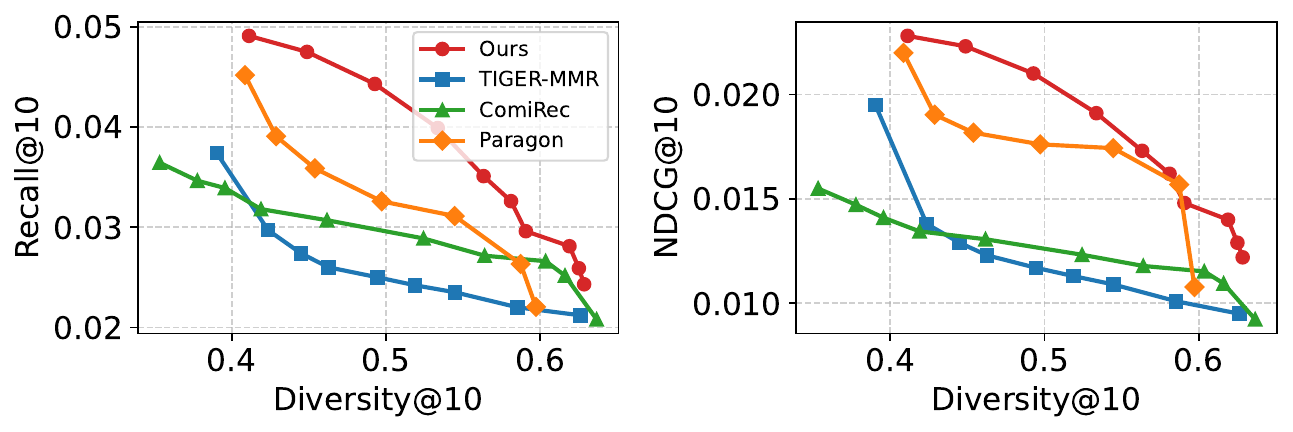}
        % \vspace{-0.6cm}
        \caption{\textbf{Scientific}} % 使用加粗
        \label{fig:pareto_scientific}
    \end{subfigure}
    
    % Main Caption: 在这里把故事讲圆
    \caption{
        Pareto frontier analysis of accuracy vs. diversity on the (a) \textbf{Instruments} and (b) \textbf{Scientific} datasets. 
        The x-axis represents Diversity@10, and the y-axes denote Recall@10 (top curves) and NDCG@10 (bottom curves). 
    }
    % \vspace{-0.45cm}
    \label{fig:pareto}
\end{figure}

\subsection{Overall Performance Comparison} 
\begin{table*}[t]
\centering

\caption{Ablation study on core components under a fixed controllability-oriented inference setting. All variants use the same controllability-oriented decoding and ranking configuration, and only the indicated component is changed. \textbf{Val} denotes the validity ratio of generated Semantic IDs.}
\label{tab:ablation}
% \vspace{0.3cm}
\small 
% 调整列间距以适应单栏宽度
% \setlength{\tabcolsep}{.9pt}

\adjustbox{max width=1.0\textwidth}{
\begin{tabular}{l ccc ccc}
\toprule
\multirow{2}{*}{\textbf{Variants}} 
& \multicolumn{3}{c}{\textbf{Instruments}} 
& \multicolumn{3}{c}{\textbf{Scientific}} \\
\cmidrule(lr){2-4} \cmidrule(lr){5-7}
& R@10 & D@10 & Val@10 
& R@10 & D@10 & Val@10 \\
\midrule
\textbf{ReGen (Full)} 
& {0.0339} & \textbf{0.5606} & \textbf{99.97\%}
& {0.0414} & \textbf{0.5051} & \textbf{99.97\%} \\
\midrule
\textit{w/o Guidance} ($\lambda=0$) 
& \textbf{0.0355} & 0.4851 & 99.97\%
& \textbf{0.0436} & 0.4644 & 99.97\% \\

\textit{w/o Refinement} ($\tau_t = 0$) 
& 0.0130 & 0.7321 & 99.97\%
& 0.0267 & 0.7307 & 99.97\% \\

\textit{w/o Topology} (No Trie) 
& 0.0136 & 0.0001 & {27.63\%}
& 0.0315 & 0.0151 & {29.76\%} \\
\bottomrule
\end{tabular}
}
% \vspace{-0.25cm}
\end{table*}
Table~\ref{tab:masked_results} summarizes standard top-$K$ accuracy under the accuracy-oriented configuration on three datasets. This table is used to verify whether the masked-refinement backbone provides a reasonable utility basis before controllability-oriented decoding is applied. Controllability itself is evaluated separately in Section~\ref{sec:Accuracy-Diversity Trade-off Analysis}. \begin{itemize}
\item \textbf{Accuracy vs. strong SOTA baselines.} Under the accuracy-oriented setting, ReGen achieves reasonable top-$K$ utility on all three datasets. On \textit{Instruments} and \textit{Scientific}, ReGen obtains Recall@10 of 0.0646 and 0.0491, respectively, ranking second among all compared methods. On \textit{Video Games}, ReGen is less competitive than the strongest accuracy-oriented baselines such as SASRec, RPG, and LLaDA-Rec. These results suggest that the masked-refinement backbone provides a usable utility basis, although ReGen is not designed to be the strongest pure-accuracy recommender. 

\item \textbf{Comparison with controllable methods.} Table~\ref{tab:masked_results} compares standard accuracy when no additional diversity constraint is emphasized. In this accuracy-oriented comparison, Paragon achieves higher Recall@10 on \textit{Instruments} and \textit{Scientific}, while ReGen remains competitive but is not always the best-performing method. The main advantage of ReGen is therefore not pure accuracy, but its ability to perform validity-preserving inference-time control, which is examined in the following accuracy--diversity trade-off analysis. \end{itemize}

\subsection{Accuracy-Diversity Trade-off Analysis}
\label{sec:Accuracy-Diversity Trade-off Analysis}

We evaluate controllability by checking how utility changes when we ask for more diverse top-10 recommendations. 
Figure~\ref{fig:pareto} plots Recall@10 and NDCG@10 against Diversity@10 on \textit{Instruments} and \textit{Scientific}. A better method keeps higher Recall/NDCG at the same diversity level.

\subsubsection{A near-plateau region on \textit{Instruments}.}
ReGen shows a clear near-plateau behavior on \textit{Instruments}. When Diversity@10 increases from 0.41 to 0.59, Recall@10 only decreases slightly from 0.0427 to 0.0394, and NDCG@10 from 0.0205 to 0.0188. In contrast, TIGER-MMR drops much faster over the same range, with Recall@10 falling from 0.0433 to 0.0319 and NDCG@10 from 0.0203 to 0.0147. At a matched diversity level around 0.54, ReGen achieves $0.0398/0.0190$ (Recall@10/NDCG@10), higher than TIGER-MMR ($0.0351/0.0163$) and ComiRec ($0.0316/0.0167$). 
This indicates that ReGen can increase diversity with a relatively small utility loss within the reported operating range.

\subsubsection{Trade-off advantage on \textit{Scientific}.} 
On \textit{Scientific}, increasing diversity is harder for all methods, but ReGen still maintains better utility at comparable diversity. Around Diversity@10 $\approx 0.56$, ReGen reaches Recall@10 $=0.0351$ and NDCG@10 $=0.0173$, compared to $0.0235/0.0109$ for TIGER-MMR and $0.0272/0.0118$ for ComiRec; Paragon also trails in Recall@10 at a similar diversity level (0.0311). At a higher diversity level around 0.59, ReGen remains the strongest in Recall@10 (0.0296), while TIGER-MMR, ComiRec, and Paragon drop to 0.0220, 0.0266, and 0.0263, respectively. 
Overall, among the controllable methods compared in Figure~\ref{fig:pareto}, ReGen retains higher Recall@10 at comparable diversity levels.

These results suggest the benefit of intrinsic guidance: ReGen guides generation directly (in logit space) rather than re-ranking a fixed candidate set, which can make it easier to keep recommendations relevant while increasing diversity.

\subsection{Ablation Study}

\begin{figure*}[t]
    \centering
    % 子图 (a): Lambda Sensitivity
    \begin{subfigure}[b]{1\textwidth}
        \centering
        % 请确保文件名与您保存的一致
        \includegraphics[width=\textwidth]{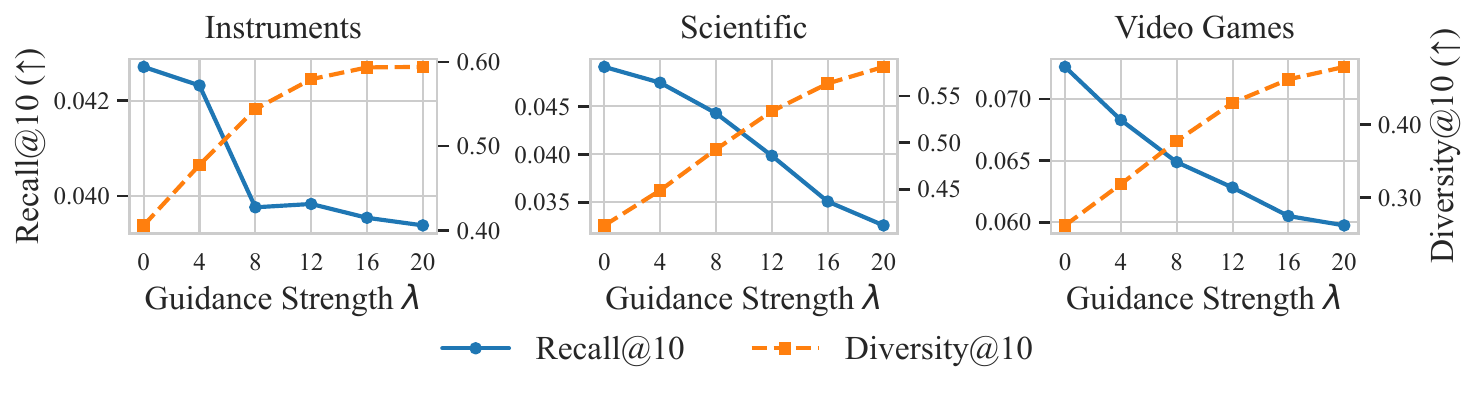}
        \caption{\textbf{Impact of Guidance Strength ($\lambda$)}: A larger $\lambda$ generally improves diversity, while overly strong guidance may reduce recall.}
        \label{fig:sens_lambda}
    \end{subfigure}
    
    % \par\bigskip % 增加垂直间距，让两张图不挤在一起
    
    % 子图 (b): Steps Sensitivity
    \begin{subfigure}[b]{1\textwidth}
        \centering
        % 请确保文件名与您保存的一致
        \includegraphics[width=\textwidth]{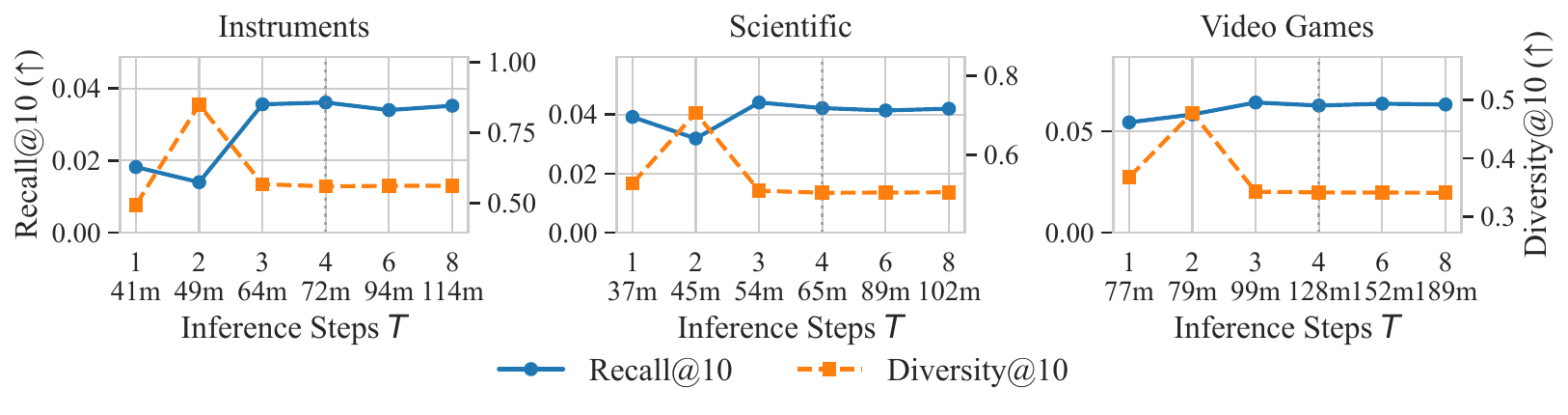} 
        \caption{\textbf{Impact of Inference Steps ($T$)}: Performance converges rapidly at $T=4$, offering the best trade-off between accuracy and efficiency.} 
        \label{fig:sens_steps}
    \end{subfigure}
    
    % Main Caption
    \caption{
       We evaluate the impact of (a) guidance strength $\lambda$ and (b) the number of inference steps $T$ on recommendation performance across the Instruments, Scientific, and Video Games datasets.
    }
    % \vspace{-0.5cm}
    \label{fig:sensitivity_analysis}
\end{figure*}

Table~\ref{tab:ablation} presents the ablation results on the \textit{Instruments} and \textit{Scientific} datasets under a fixed controllability-oriented setting. The purpose is to compare the relative contribution of each component under the same decoding and ranking configuration.

\begin{itemize}
    \item \textbf{Guidance drives controllability.} 
    Turning off guidance ($\lambda=0$) removes the logit-level control signal while keeping the same decoding and ranking pipeline. As shown in Table~\ref{tab:ablation}, while R@10 increases marginally (e.g., $0.0339 \to 0.0355$ on Instruments), D@10 drops significantly ($0.5606 \to 0.4851$). This shows that the guidance signal is the key factor that guides generation toward more diverse recommendations at inference time.

    \item \textbf{Refinement preserves utility under control.} Removing confidence-aware refinement ($\tau_t=0$) leads to a sharp accuracy drop (e.g., \textit{Instruments}: R@10 $0.0339 \rightarrow 0.0130$), accompanied by a spike in diversity to $\approx 0.73$. We identify this as \textbf{illusory diversity}: without iterative correction, strong guidance disrupts the semantic coherence of the sequence, yielding random (diverse) but irrelevant items. The refinement module is critical for maintaining recommendation quality when applying control.

    \item \textbf{Topology constraint ensures valid generation.} 
    Removing Topology-Constrained Tree causes a severe validity collapse (Val@10 drops to $27.63\%$ on \textit{Instruments} and $29.76\%$ on \textit{Scientific}). Since the majority of generated codes cannot be mapped to real items, recommendation utility inevitably suffers a sharp decline. This indicates that the semantic-ID space is sparse, and the topology constraint is necessary to keep generation within the set of valid items.

\end{itemize}

\subsection{Parameter Sensitivity and Efficiency}

We evaluate the robustness of \textbf{ReGen} regarding two key inference parameters: the guidance strength $\lambda$ and the refinement iterations $T$. Results are summarized in Figure~\ref{fig:sensitivity_analysis}.

\begin{itemize}
\item \textbf{Impact of Guidance Strength ($\lambda$).}
The parameter $\lambda$ controls the intensity of the logit-space intervention. As shown in Figure~\ref{fig:sensitivity_analysis}(a), increasing $\lambda$ consistently improves Diversity@10. 
Crucially, ReGen maintains high accuracy within a broad practical range (e.g., $\lambda \in [4,12]$), exhibiting a stable accuracy--diversity trade-off. 
This robustness stems from our hierarchical design: guidance primarily guides coarse-grained semantic planning, while fine-grained details are subsequently repaired by confidence-aware refinement. Performance degradation occurs only under excessively strong guidance ($\lambda > 20$), which overwhelms the refinement process.

\item \textbf{Refinement Steps ($T$) and Efficiency.}
We analyze the impact of iterations $T$ on both convergence and total evaluation time, which is annotated on the x-axis in Figure~\ref{fig:sensitivity_analysis}(b).
The model exhibits rapid convergence, with metrics stabilizing at around $T=4$. 
Early steps ($T<3$) reflect an exploratory phase with higher diversity variance, while subsequent steps effectively fix low-confidence tokens.
Since inference cost increases with $T$, we identify $T=4$ as a practical operating point that achieves stable performance with lower evaluation cost than larger $T$ values. 
These results confirm that ReGen balances controllability and efficiency without requiring delicate hyperparameter tuning.

\end{itemize}

\section{Limitations and Future Work.}
This work focuses on diversity control as a representative case of inference-time controllable generative recommendation. Although the proposed logit guidance can in principle be instantiated with other control signals, such as fairness, exposure balancing, or freshness, we leave a systematic evaluation of these objectives to future work. In addition, ReGen relies on the quality of the underlying Semantic ID tokenizer and its hierarchical organization. If the learned codebooks are noisy or poorly aligned with item semantics, coarse-level guidance may become less effective. 
Future work will study tokenizer-agnostic control mechanisms. Furthermore, beyond category coverage, richer notions of controllability such as novelty, serendipity, calibration, fairness, and exposure balancing require objective-specific control functions and evaluation protocols, which we also leave for future work.

\section{Conclusion}

In this work, we propose ReGen, a framework for real-time controllable generative recommendation with training-free inference-time control. Our method employs parallel iterative denoising to enable flexible control without parameter updates, thereby addressing the rigidity and instability of existing generative recommenders when discrete control signals are introduced.
ReGen combines logit-space guidance for coarse-grained semantic control with confidence-aware refinement to iteratively repair low-confidence tokens and preserve relevance and validity. Extensive experiments show that ReGen achieves a favorable accuracy–diversity trade-off and converges efficiently with only a few refinement steps.

% \balance
% \bibliographystyle{elsarticle-num}
\bibliographystyle{IEEEtran}
\bibliography{conference}

\end{document}